\documentclass[structabstract]{aa} 
\usepackage{txfonts}
\usepackage{graphicx}
\usepackage{natbib}
\bibpunct{(}{)}{;}{a}{}{,} 

\newcommand{\dx}{\mathrm{d}}
\def\phx{{\texttt{PHOENIX}}}

\begin{document}

\title{Time-dependent radiative transfer with \phx }
\author{D. Jack\inst{1}
  \and P. H. Hauschildt\inst{1} 
   \and E. Baron\inst{1,2}} 

\institute{Hamburger Sternwarte, Gojenbergsweg 112, 21029 Hamburg, Germany\\
  e-mail: djack@hs.uni-hamburg.de; yeti@hs.uni-hamburg.de
  \and Homer L. Dodge Department of Physics and Astronomy, University of Oklahoma, 440 W Brooks, Rm 100, Norman, OK 73019-2061 USA\\
  e-mail: baron@ou.edu}

\date{Received 8 September 2008 /
      Accepted 3 April 2009}

\abstract {} 
{We present first results and tests of a time-dependent extension to
the general purpose model atmosphere code \phx. We aim to produce light curves and
spectra of hydro models for all types of supernovae.} 
{We extend our
model atmosphere code {\phx} to solve time-dependent non-grey, NLTE,
radiative transfer in a special relativistic framework.  A simple
hydrodynamics solver was implemented to keep track of the energy
conservation of the atmosphere during free expansion.}  
{The correct operation of the new additions to {\phx} were verified in test
calculations.}  
{We have shown the correct operation of our extension
to time-dependent radiative transfer and will be able to calculate
supernova light curves and spectra in future work.}


\maketitle

\section{Introduction}

All types of supernovae are important for the role that they play in
understanding stellar evolution and galactic nucleosynthesis and as
cosmological probes.
Type Ia supernovae are of particular cosmological interest, e.g., because the 
dark energy was discovered with Type Ia supernovae
\citep{riess_scoop98,perletal99}. 

In dark energy studies, the goal now is to characterize the nature of
the dark energy as a function of redshift. While there are other
probes that will be used (gravitational lensing, baryon acoustic
oscillations), a JDEM or Euclid mission will likely consider supernovae
in some form. In planning for future dark energy studies both from space
and from the ground, it is important to know whether the mission will
require spectroscopy of modest resolution, or whether pure imaging or
grism spectroscopy will be adequate. Several purely spectral
indicators of peak luminosity have been
proposed
\citep{nugseq95,hach06,bongard06a,bronder08,ffj08,ledu09}.
What is
required is an empirical and theoretical comparison of both light curve shape luminosity indicators
\citep{pskovskii77,philm15,rpk96,goldhetal01} and spectral indicators.

To make this comparison one needs to know more about the
physics going on in a supernova explosion and to be able to calculate
light curves and spectra self-consistently. 
Thus, we need to extend our code to time-dependent problems.
While our primary focus is on Type Ia supernovae, the time-dependent 
radiative transfer code is
applicable to all types of 
supernovae, as well as to other objects, e.g., stellar pulsations.

In the following we present the methods we used to implement time
dependence. First we focus on solving the energy equation (first law)
and then turn our attention to time-dependent radiative transfer.

\section{Equation of energy conservation}

To compute light curves we extended \phx\ with a
time-dependent solution of the non-grey, NLTE radiative transfer
problem in special relativistic environments and a simple hydrodynamic
code to solve the free expanding case relevant to supernova light
curves.  The idea is to keep track of the conservation of energy for
the gas and radiation together and to allow for different time scales
for the gas and the radiation.

For the time-dependent radiative transfer problem we extended our
existing radiative transfer code \phx\ \citep{hbjcam99,hbmathgesel04}.
The change in the energy density of a radiating material is given by
equation (96.15) in \citet[]{found84}
\begin{equation} \frac{D}{Dt}E=-\frac{\partial}{\partial
M_{r}}\left(L_{r}+P_{r}\right)+\varepsilon,\label{eqn:energy}
\end{equation} 
where $E$ is the total energy density. All quantities
are considered in the comoving frame. $P_{r}$ is not the pressure, but rather
mechanical power on the sphere of a radius $r$. 
Equation~\ref{eqn:energy} is only valid to first order in v/c, and thus lacks
the full special relativistic accuracy of \phx. 
This is adequate for the velocities found in supernovae.
The total energy density of a
radiating fluid consists of the sum of the energy density of the
material, the energy density of the radiation field, the kinetic
energy density of the material, and the gravitational energy density:
\begin{equation} 
E=E_{gas}+\frac{E_{0}}{\rho}+E_{kin}+E_{grav}.
\end{equation} 
For supernovae in the free expansion phase, the
gravitational energy density $E_{grav}$ is negligible since the potential 
is small in absolute value with the standard choice of zero at infinity. 
Homologous expansion is a reasonably good assumption for supernovae.
The energy release by the decay of $^{56}$Ni
can influence
the dynamics of the expansion \citep{pinto00}. \citet{woosley07} 
compared a study following this energy release to the results from assuming
homologous expansion. Figure 2 in their paper shows the deviation  
and density variations can be as large 
as 10\%. However, this is probably an upper limit 
due to the simple burning parameterization used in that study. 
Ultimately, when the deflagration to detonation
transition is understood, it will be important to revisit this issue
and replace 1-D calculations with full 3-D calculations, which include
the effects of clumps, as well as nickel bubble expansion. 
For now the accuracy of homologous expansion should be adequate, given
the other uncertainties in the problem.

With the assumption of homology, the velocity of a given matter
element is then constant as   
is the kinetic energy
density. Thus, we can neglect the kinetic energy term $\frac{D
E_{kin}}{Dt}$. So for our approach, we only have to consider the energy
densities of the radiation field and the material. For the material,
this includes effects such as an energy deposition due to radioactive
decay of nickel and cobalt in an SN Ia envelope.

The other possible cause of a change in the energy density is the
structure term. This term is given by \citep{cvb1}
\begin{equation} 
\frac{\partial}{\partial
M_{r}}\left(P_{r}+L_{r}\right)=\frac{\partial}{\partial
M_{r}}\left\{4\pi
r^{2}\left[u\left(p+P_{0}\right)+F_{0}\right]\right\}
\end{equation} 
where $p$ is the pressure of the material and $P_{0}$
the radiation pressure, $u$ the velocity of the expanding gas,
the radiative flux is represented by $F_{0}$, and the mass inside of
the radius $r$ of a layer is given by $M_{r}$. The radiation pressure
is a result of the solution of the detailed radiative transfer equation 
and given by
\begin{equation}
P_{0}=\frac{4\pi}{c}K,
\end{equation}
with $K$ the second moment of the radiation field.
The change of the energy density is given by the two quantities

\begin{equation}
L_{r}=4\pi r^{2}F_{0}
\end{equation}
and
\begin{equation}
P_{r}=4\pi r^{2}u\left(p+P_{0}\right).
\end{equation}
If the atmosphere is
in radiative equilibrium, the structure term is zero and the energy
density stays constant if there is no additional energy source and the
atmosphere is not expanding.

All the quantities required for the structure term can be derived from
thermodynamics or the solution of the radiative transfer problem.  We
need the energy density of the material $E_{gas}$ and the energy density of
the radiation field $\frac{E_{0}}{\rho}$. For the latter, we have to
solve the radiative transfer equation for the radiation field and the
radiative energy. We use our radiative transfer code \phx\ to solve 
the time-independent radiative transfer equation.
The energy of the radiation field is given by
\begin{equation} 
E_{0}=\frac{4\pi}{c}J,
\end{equation} 
where $J$ is the mean intensity and $c$ the speed of
light.

The energy density of the material is given by
\begin{equation} 
E_{gas}=\frac{3}{2}\frac{p}{\rho}=\frac{3}{2}\frac{R}{m_{u}}T,
\end{equation}
with the mean molecular weigh $m_{u}$ and the universal
gas constant $R$. The gas pressure is represented by $p$ and the
density by $\rho$. $T$ stands for the temperature of the gas.
The sum of the radiation and material energy density 
is then the total energy density

\begin{equation}
E_{total}=E_{gas}+\frac{E_{0}}{\rho}.
\label{eq:totenergy}
\end{equation}

The change in this total energy density is given by Eq. 1.
So the equation to calculate the new energy density $E_{new}$ is given by

\begin{equation}
E_{new}=E_{old}
-\frac{\partial}{\partial
M_{r}}\left(L_{r}+P_{r}\right)\Delta t+\varepsilon\Delta t.
\label{eq:dedt}
\end{equation}

We now have all the needed equations to calculate a simple light curve.  
One problem for the calculation is that we can only determine
the change in the {\sl total} energy density for the next time step.
However, the total energy change is divided into a change in the gas
energy density and the energy density of the radiation field.
To obtain the correct distribution of the gas and radiative
energy, one has to iterate for each time step by solving the radiative
transfer equation to compute the correct new temperature at the next
time step. 

To get the correct new temperature we apply the following
iteration scheme.
The error in the energy density, $E_{err}$ is given by 
\begin{equation}
E_{err}=\frac{E_{current}-E_{target}}{E_{target}}.
\end{equation} 
Here, $E_{target}$ is the desired new total energy density, 
which is known from Eq. \ref{eq:dedt},
and $E_{current}$ is the total energy density obtained by 
Eq. \ref{eq:totenergy} with the current
temperature guess and the radiative transfer solution. Tests have shown
that the error is almost linearly proportional to the temperature
$T$. Therefore, a new temperature
guess can be calculated for the next iteration step. 
The new temperature guess $T_{new}$ is obtained by
\begin{equation}
T_{new}=\frac{E_{err} T_{old} - E_{err_{old}}T_{cur}}{E_{err}-E_{err_{old}}},
\end{equation}
where $T_{cur}$ and $E_{err}$ are the current temperature guess and energy error. 
The variables $T_{old}$ and $E_{err_{old}}$ are the temperature and energy error 
of the temperature iteration step before.
With the new temperature guess we solve the radiative transfer equation
again and check whether the total energy density is the desired one.
It takes approximately four or 
five iteration steps to get the correct new temperature for a typical 
time step. 
The energy density is correct within an accuracy of $10^{-5}$.

\section{Test light curves}

For the first test calculations we have solved the radiative transfer
equation for a grey test atmosphere. Test model atmospheres are
divided in 100 layers and we consider here time independent radiative
transfer (in \S~\ref{sec:tdrt} below we discuss time dependent radiative transfer).

As a first test we applied our time evolution code to a static
atmosphere. The test atmosphere is not expanding and no energy
sources are present. Inside the test atmosphere we used a
``lightbulb'' radiating with a constant luminosity to simulate the
internal energy flow from a star. We assumed an approximate
temperature structure for $t=0$ and then let the atmosphere evolve
towards radiative equilibrium. The atmosphere changed until it
reaches steady state and the luminosity is constant in both space and time.
The resulting final temperature structure should be identical to the
structure for radiative equilibrium computed directly.

\begin{figure}
\resizebox{\hsize}{!}{\includegraphics{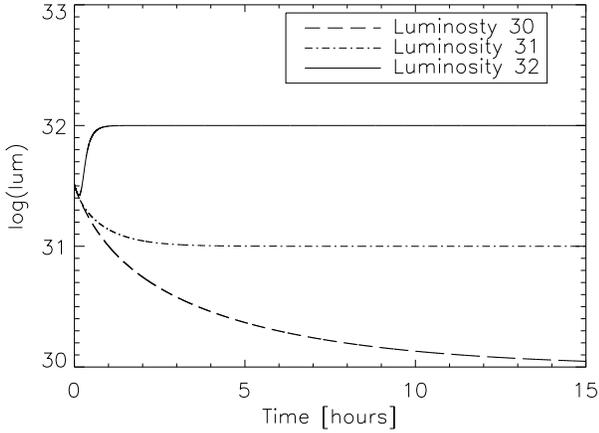}}
  \caption{Three different light curves for the evolution to a
stationary state. The three models have different
luminosities produced by different inner ``lightbulbs''.}
  \label{fig:lc}
\end{figure}

In Fig. \ref{fig:lc} we show the light curves for three different
static models. We can observe the model on its way towards radiative
equilibrium. All calculations were started from the same temperature
structure.  After a certain time, the radiative relaxation time scale,
each atmosphere has the (constant) luminosity of the lightbulb
throughout the configuration.

\begin{figure}
\resizebox{\hsize}{!}{\includegraphics{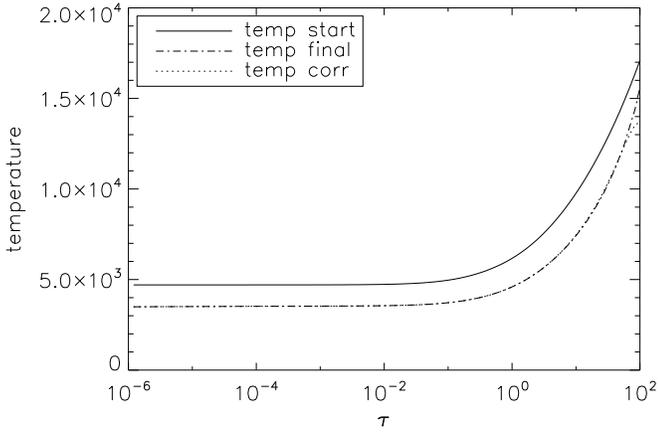}}
  \caption{The final temperature structures of the time evolution
calculations.} 
  \label{fig:tcor}
\end{figure}

The final temperature structure of the model atmosphere with the
lightbulb with a luminosity of $10^{31}\,$erg/s is displayed in
Fig. \ref{fig:tcor}. Also shown is the result of a calculation using
the temperature correction procedure of {\phx} \citep{phhtc03} to
directly compute the radiative equilibrium structure of the
configuration. As one can clearly see, the resulting temperature
structure of both methods agrees very well, and the maximum
deviation of the temperature structure is less than one percent. Only
the inner three layers have a deviation of up to 10 percent due to 
the implicit assumption of a diffusion approximation (and not a lightbulb)
in the static calculation.

\begin{figure}
\resizebox{\hsize}{!}{\includegraphics{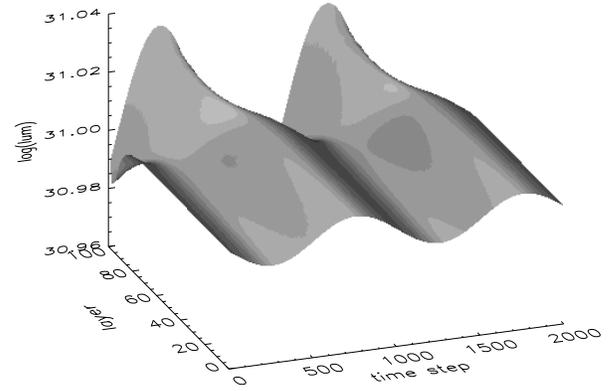}}
  \caption{ The flux in every layer at each time step.}
  \label{fig:sinus_surf}
\end{figure}

The next test is to look at time varying atmospheres. As an example we
considered an atmosphere with a sinusoidally varying lightbulb
inside.  The resulting luminosity in each layer at each time point is
plotted in Fig. \ref{fig:sinus_surf}.  The luminosity in each layer is
sinusoidal. It takes some time for the radiation to reach the outside
boundary of the atmosphere and this results in a phase shift compared
to the lightbulb.

\begin{figure} 
\resizebox{\hsize}{!}{\includegraphics{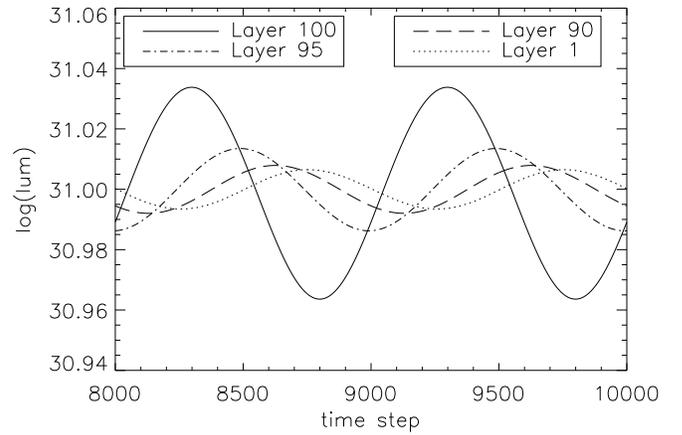}}
  \caption{The result for a sinusoidal varying lightbulb.
Shown here is the luminosity in different layers. The phase
shift between the lightbulb and the emergent flux is roughly $\pi$.}
  \label{fig:sinus}
\end{figure}

We now take a look at the luminosity in different layers. These are
plotted in Fig. \ref{fig:sinus}. One can see the phase shift of the sine
in each layer. As expected, the luminosity from the central source
needs time to move through the atmosphere.

One can also see that the amplitude of the sine decreases with
increasing radius. What one would expect is that the amplitude is the
same in every layer. The radiation is moving outwards and the incoming
varying luminosity from the lightbulb should move through every layer
and therefore the amplitude is supposed to be the same in each layer.
If we integrate the luminosity over a whole sine, it stays constant in
every layer. In the plot one can see that the mean luminosity is at
the same level, so the luminosity is preserved. But why does the
amplitude decrease? A possible explanation is that the radiation
moving through the atmosphere is going backwards when the
sine is declining, so this smears out the amplitude. If the sine
varying radiation is moving through a very thick atmosphere, the
amplitude at the top of the atmosphere is finally flat. An observer
sees the mean luminosity.

\begin{figure}
\resizebox{\hsize}{!}{\includegraphics{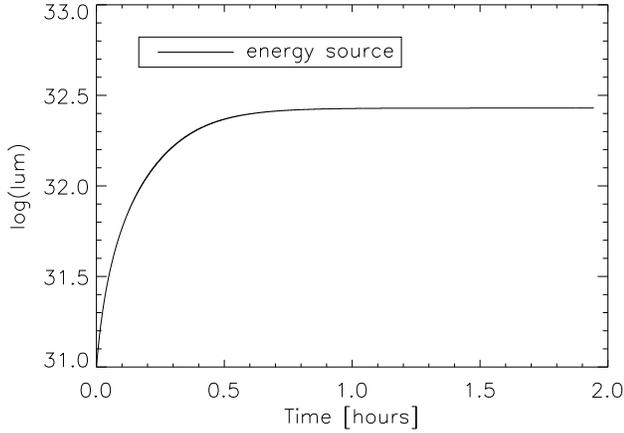}}
  \caption{The light curve of an lightbulb with an additional energy source.
An energy source in each layer causes an increasing
luminosity of the model atmosphere.}
  \label{fig:source}
\end{figure}

For the next test we consider an atmosphere with an internal energy
source. The initial structure is the radiative equilibrium structure of
the static model with the lightbulb with a luminosity of
$10^{31}\,$erg/s.  We assume a constant energy deposition rate in
each layer of the model atmosphere. The luminosity is expected to
increase over time and towards the outside. Figure \ref{fig:source}
shows a plot of the light curve of this test atmosphere. 
The luminosity increases in time because of the energy deposition 

\subsection{Expansion}

In the case of SN Ia, homologous expansion is a good assumption after
the initial breakout.  In our test model each layer has a
constant velocity, $u\propto r$, to simulate a freely expanding
envelope.  The new radius $r_{new}$ of a layer is, therefore,
determined by
\begin{equation} r_{new}=u\cdot\Delta t+r_{old}
\end{equation} for a time step $\Delta t$, while the layer is
expanding with a velocity $u$.  The radius before the new time step is
$r_{old}$. With the same assumption of homologous expansion the new
density $\rho_{new}$ of a layer after at the new time step is
determined by

\begin{equation}
\rho_{new}=\rho_{old}\cdot\left(\frac{r_{old}}{r_{new}}\right)^{3}
\end{equation}
where $\rho_{old}$ is the old density. With the new radius and
density, we are now able to calculate the new thermodynamics of the
atmosphere. 
We assume the expansion of the
supernova is an adiabatic process. The internal energy of the material
changes due to work $\dx W$ done during the expansion
\begin{equation} 
\dx E =\dx W = - p \dx V,
\end{equation} 
where $p$ is the pressure and $\dx V$ a volume change.
The energy conservation equation considers the energy density. The
change in the energy density for the adiabatic expansion is given by
\begin{equation} 
\dx E_{adia} = \frac{\dx W}{m}= - \frac{p}{m} \dx V.
\end{equation} 
The mass $m$ of a layer is given by its volume and density
\begin{equation} 
m=V\cdot \rho.
\end{equation} 
In our approach to a homologous expanding supernova of
the type Ia, we consider each layer has a constant mass, so the
derivative $\dx V$ is given by
\begin{equation} 
\dx V = -\frac{m}{\rho^{2}} \dx \rho.
\end{equation} 
Together with the equation of state
\begin{equation}
 p=\frac{R}{\mu}T\rho,
\end{equation}
we obtain the work for the adiabatic expansion as
\begin{equation} 
\frac{\dx W}{m}=\frac{R}{\mu}T\frac{\dx \rho}{\rho}.
\end{equation} 
For differences in a time step $\Delta t$, the work is
\begin{equation} 
W_{adia}= \Delta \frac{W}{m}=\frac{R}{\mu}T \ln
\left( \frac{\rho_{2}}{\rho_{1}} \right).
\end{equation}
 For the first test of a light curve for a supernova
with an expanding atmosphere, we neglect the interaction between the
layers, meaning the structure term is equal to the work of the
adiabatic expansion so that
\begin{equation}
\frac{D}{Dt}E=\frac{D}{Dt}\left(\frac{E_{0}}{\rho}+e\right)=W_{adia}.
\end{equation}

As we now consider expanding atmospheres, we use more supernova-like
structures for the tests.  We set the maximum velocity to 30000 km/s
and use a radius of $5\times 10^{15}\,$cm.  With this setup, we
calculated a light curve for the expanding atmosphere.

\begin{figure}
\resizebox{\hsize}{!}{\includegraphics{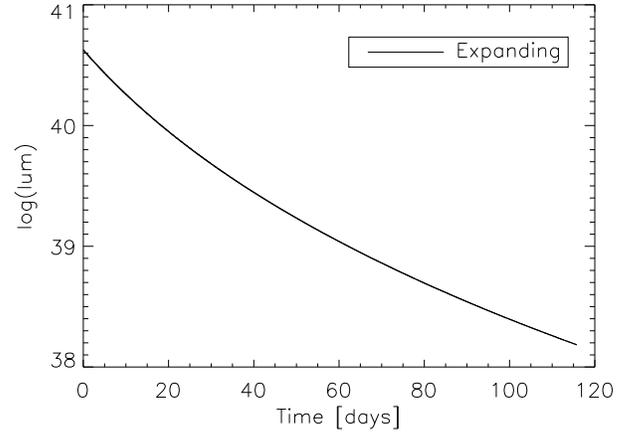}}
  \caption{Light curve of an atmosphere that is just expanding.}
  \label{fig:expan}
\end{figure}

Figure \ref{fig:expan} shows a plot of the light curve of supernova test
atmosphere that is simply expanding. We considered time-independent
radiative transfer and a grey atmosphere. As one can see, the
luminosity is decreasing, because the atmosphere is cooling down
adiabatically.

\begin{figure}
\resizebox{\hsize}{!}{\includegraphics{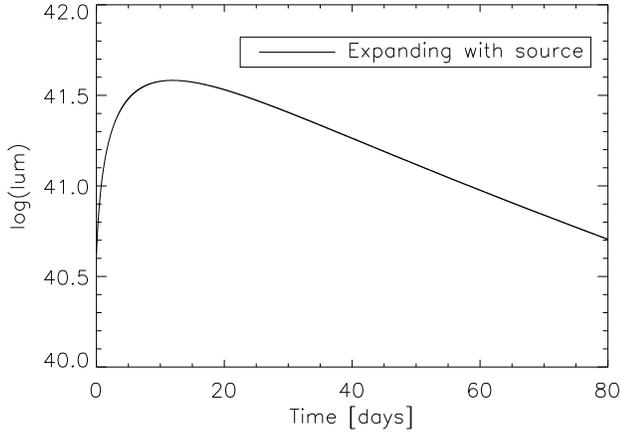}}
  \caption{Light curve of an atmosphere that is expanding and has an
energy source. It has the typical shape of a light curve of a SN
Ia. The luminosity rises due to the energy deposition. After the
maximum we see the decline resulting from the expansion.}
  \label{fig:expansource}
\end{figure}

Now we test a setup more closely resembling a real
supernova light curve. Therefore, we take an initial atmosphere
structure and add an energy source (radioactive decay) in each
layer. The energy source exponentially decreases to simulate
declining activity of the radioactive species. Figure
\ref{fig:expansource} shows the plot of the light curve of this test,
resulting in a light curve with a supernova-like shape. We have a
rising part of the light curve at the beginning because of the energy
deposition. After the maximum, the luminosity decreases due to the
ongoing expansion and decreasing energy deposition.  Of course this
light curve is far from correct because the assumption of a grey
atmosphere is a bad assumption for an SN Ia. But the tests show that
the code behaves as expected.

\subsection{Entropy} 

We calculated the entropy to test the code. 
Because energy is moving through the
atmosphere, entropy is not conserved.  
Nevertheless, testing entropy conservation is an
important test of the correctness of the code. Therefore,
we calculated the entropy change in the case of pure 
adiabatic expansion. 
For testing, we neglected interaction between layers and
gamma ray deposition and calculated the entropy change 
for a time step for the just expanding case.

To be consistent with our hydrodynamics equations, we deduced the entropy from the first law of 
thermodynamics. 
A change in the entropy during a time step is therefore
given by
\begin{equation} \frac{\Delta S}{mR}=\frac{3}{2}\frac{1}{\mu}\ln\left(
\frac{T_{2}}{T_{1}}\right)+
\frac{3}{2}\left(\frac{1}{\mu_{2}}-\frac{1}{\mu_{1}}\right)-\frac{1}{\mu}\ln\left(\frac{\rho_{2}}{\rho_{1}}\right)
\end{equation} where 2 is the index of quantities at the new time and
1 the one of the old. For the integration of the temperature and the density
term, we kept $\mu$ fixed. This is not correct, but it is simpler to solve and the 
resulting differences for the entropy are small.

The entropy given here is only the entropy of the gas, so to
check that the expansion worked right we had to neglect the radiation
energy density in our energy density conservation equation. 
Furthermore we ignored the interaction between the layers and neglected the
structure term.  We now have only a change in the energy due to the
adiabatic expansion.

With this setup we calculated a few time steps and checked the
entropy. Even for a long time step of 1000s the entropy stays almost constant.
The relative change of the entropy was  $\approx 10^{-6}$.

\section{Time-dependent radiative transfer\label{sec:tdrt}}

So far, our radiative transfer code only solves the time-independent
radiative transfer equation. For an implementation of the time
dependence in the radiative transfer itself, the spherical symmetric
special relativistic radiative transfer equation (SSRTE) for expanding
atmospheres \citep{hbjcam99} is extended so that the additional time
dependent term is given by
\begin{equation} \frac{\gamma}{c}\left(1+\beta\mu\right)\frac{\partial
I}{\partial t}
\end{equation}
where $\beta=\frac{v}{c}$ is the velocity in units of
the speed of light $c$ and $\gamma=(1-\beta^{2})^{-1/2}$ is the usual
Lorentz factor. Here, $I$ is the intensity, $\mu$ the cosine of the
angle between the radial direction and the propagation vector of the
light. Using the notation of \citet{hb04}, the comoving frame SSRTE with the
additional time dependent term is given by
\begin{equation} 
a_{t}\frac{\partial I}{\partial t}
+a_{r}\frac{\partial I}{\partial r}+a_{\mu}\frac{\partial I}{\partial
\mu}+a_{\lambda}\frac{\partial \lambda I}{\partial \lambda}
+4a_{\lambda}I=\eta-\chi I
\end{equation} 
where $\eta$ is the emissivity and $\chi$ the extinction
coefficient. The wavelength is represented by $\lambda$.  The
additional time dependent coefficient is given by
\begin{equation} 
a_{t}=\frac{\gamma}{c}\left(1+\beta\mu\right).
\end{equation} 
Along the characteristics the equation has the form
\begin{equation} \frac{\dx I_{l}}{\dx s}+a_{t}\frac{\partial
I}{\partial t}+a_{l}\frac{\partial \lambda I}{\partial \lambda}
=\eta_{l}-(\chi_{l}+4a_{l})I_{l}
\end{equation}
where $\dx s$ is a line element along a (curved) characteristic
and $I_{l}()$ is the specific intensity along the characteristic at point
$s \ge 0 $($s=0$ denotes the beginning of the characteristic). For a
definition of the other coefficients see \citet{hb04}.

\subsection{Discretization of the time derivative}

We used the first discretization as described in \citet{hb04} and added
the time dependent term. We discretized the time-dependent, as well as
the wavelength, derivative in the SSRTE with an fully implicit method.
The discretization of the time dependent term is given by
\begin{equation} 
\left.\frac{\partial I}{\partial t}\right|_{t=t_{j}}
=\frac{I_{t_{j}}-I_{t_{j-1}}} {t_{j}-t_{j-1}},
\end{equation} 
so the SSRTE including the time discretization can
be written as
\begin{equation} 
\frac{\dx I}{\dx
s}+a_{\lambda}\frac{\lambda_{l}I-\lambda_{l-1}I_{\lambda_{l-1}}}{\lambda_{l}-\lambda_{l-1}}
+a_{t}\frac{I-I_{t_{j-1}}}{t_{j}-t_{j-1}}=\eta_{\lambda_{l}}-(\chi_{\lambda_{l}}+4a_{\lambda})I,
\end{equation}
where $I$ is the intensity at wavelength point $\lambda_l$ and time
point $t_j$. We define the optical depth scale along the ray as
\begin{equation} 
\dx\tau=\chi+a_{\lambda}\left(4+\frac{\lambda_{l}}{\lambda_{l}-\lambda_{l-1}}\right)+\frac{a_{t}}{\Delta
t}=-\hat{\chi}\dx s.
\end{equation} 
Introducing the source function $S=\eta/\chi$, we get
\begin{equation} 
\frac{\dx
I}{\dx\tau}=I-\frac{\chi}{\hat{\chi}}\left(S+
\frac{a_{\lambda}}{\chi}\frac{\lambda_{l-1}}{\lambda_{l}-\lambda_{l-1}}I_{\lambda_{l-1}}
+\frac{a_{t}}{\chi}\frac{1}{\Delta t}I_{t_{j-1}} \right) \equiv
I-\hat{S}.
\end{equation} 
We now have a modification of the source function and a
new definition of the optical depth scale along a ray. With this
redefinition of the optical depth and the source function, one can now
proceed with the formal solution as described in \citet{hb04}.

\subsection{Test calculations: time-dependent radiative transfer}

For the test of the time-dependent radiative transfer, we used a static
atmosphere structure to see the direct effects of the time dependence
of the radiative transfer. Therefore, the temperature, radius, and
density are all constant in time. For the test we then changed the
inner boundary condition for the radiation (the ``lightbulb'') to
initiate a perturbation of the radiation field, which then moves
through the atmosphere via the time-dependent radiation transfer.

For the first test we switched on an additional light source inside
the atmosphere. This light source has a luminosity of $10^{9}$ times
higher then the inner lightbulb.
\begin{figure}
\resizebox{\hsize}{!}{\includegraphics{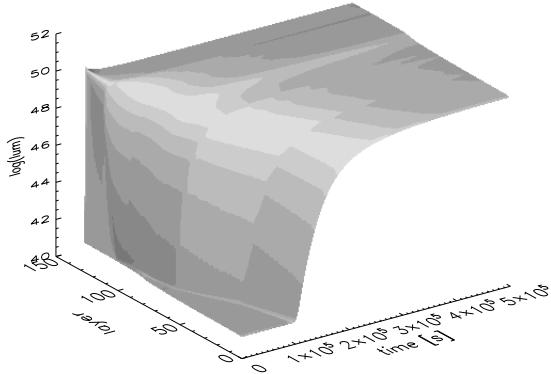}}
  \caption{Luminosity of each layer and time step of our test atmosphere.}
  \label{fig:e9_tdep}
\end{figure} In Fig. \ref{fig:e9_tdep} we show the result of the
time-dependent radiative transfer calculation. One can see that the light
propagates outwards through the atmosphere. It takes a while before
the additional light has propagated everywhere
throughout the atmosphere. In 
\begin{figure}
\resizebox{\hsize}{!}{\includegraphics{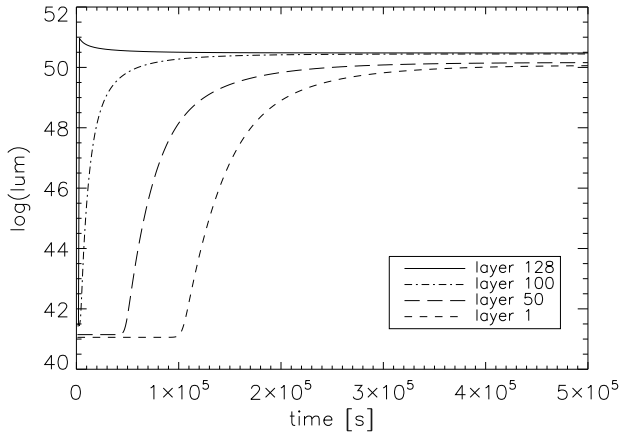}}
  \caption{The light curves of a few layers. In this case the atmosphere has 
an additional light source switching on inside. The radiation needs time to
get to the outer layers.}
  \label{fig:e9_lay_tdep}
\end{figure} 
Fig. \ref{fig:e9_lay_tdep} the light curves of a few
layers are shown. One can see that the radiation needs time to get to
the outer layers.

We compared our time scale to the radiation diffusion time scale.
Assuming a random-walk process the mean free path for a photon is given by
$\lambda_{p}=\frac{1}{\bar{\chi}}$,
where $\bar{\chi}$ is the mean opacity. 
For a travel distance $l$ the time $t_{p}$ a photon needs is given by
\begin{equation}
t_{p}\approx \frac{1}{3}\frac{l^{2}}{c\lambda_{p}}=\frac{1}{3}\frac{l^{2}}{c}\bar{\chi},
\end{equation}
where $c$ is the speed of light. For the mean opacity $\bar{\chi}$ we
used the Rosseland mean. 
In our test case the atmosphere was divided into 128 layers. The mean opacity $\bar{\chi}$
ranges from $3\cdot10^{-19}$~cm$^{-1}$ in the outer parts to 
$3\cdot10^{-14}$~cm$^{-1}$ in the inner parts of the atmosphere. The distance $l$ is 
the thickness of each layer, and it ranges between $6\cdot10^{12}$~cm and $3\cdot10^{13}$~cm.
The result is that the diffusion time for a
photon through the whole model atmosphere is about $3\cdot10^{3}$~s. 
As one can
see in our plots our time scale is roughly $10^{5}$. The assumption of a
diffusion through the atmosphere is only valid for optically thick
regions. Another problem is the choice of the right mean opacity.
Considering this our estimate of the time scale is reasonable.

An important test is also to check that the results of the 
time-dependent radiative transfer calculation depend on the size of the
time step. We tested this with a model that has a small perturbation
inside that is moving outwards. This was calculated with two
different time steps.
\begin{figure}
\resizebox{\hsize}{!}{\includegraphics{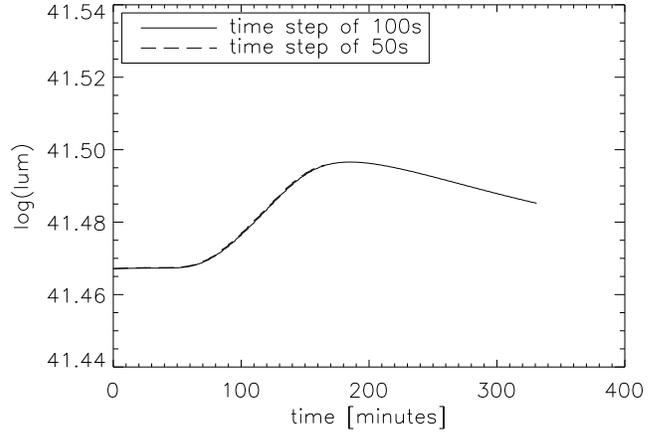}}
  \caption{The light curve of an atmosphere where a small perturbation moves outwards.  
The two plotted light curves of layer 110 were calculated with different time steps.}
  \label{fig:diff_tdep}
\end{figure} 
In Fig. \ref{fig:diff_tdep} the results of calculations
with two different time steps are shown. As one can see, the result
does not depend on the size of the time step.

In the next test we put a sinusoidally varying light bulb inside of
our test atmosphere.
\begin{figure}
\resizebox{\hsize}{!}{\includegraphics{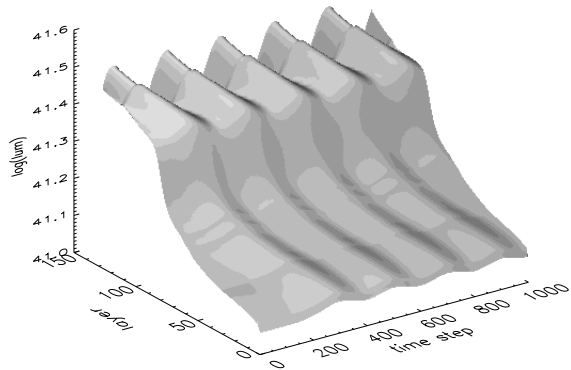}}
  \caption{Luminosity in each layer and time step. 
The light source inside the atmosphere varies sinusoidally.}
  \label{fig:sinus_tdep}
\end{figure} 
In Fig. \ref{fig:sinus_tdep}, we show a plot of the
luminosity for each time step. After a while, the luminosity of the
whole atmosphere varies sinusoidally and steady state is reached.
\begin{figure}
\resizebox{\hsize}{!}{\includegraphics{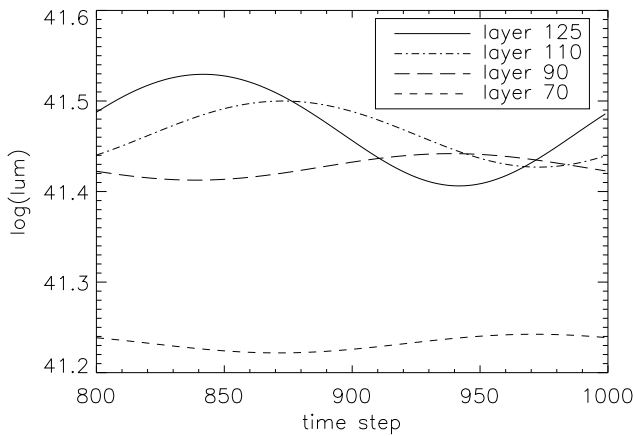}}
  \caption{Flux at different radii of a sinusoidally varying atmosphere.}
  \label{fig:sinus_lay_tdep}
\end{figure} 
The luminosities in different layers are shown in
Fig. \ref{fig:sinus_lay_tdep}. One can see a phase shift of the sine curve,
due to the time required for the radiation field to propagate through the 
atmosphere.

\section{Conclusion}

We have implemented a time evolution code into our general-purpose model
atmosphere 
code \phx, which keeps track of the energy conservation. Because a
homologous expansion is a good assumption for supernovae in general
and particularly for Type 
Ia supernovae, we considered adiabatic free expansion for our code.
With first test calculations, we reproduced the expected behavior
of the test cases. Static atmospheres adopted the luminosity of an internal
lightbulb and perturbations of the lightbulb moved outwards in time.
We also calculated a light curve that had the shape of a typical
supernovae Ia light curve.

To complete the physics of time-dependent processes in a supernova
atmosphere, we extended our code even further. \phx\ now solves the time
dependent radiative transfer equation. We presented the new
discretization scheme of the time dependent $\frac{\partial}{\partial
t}$ term. We checked our code with a series of tests. A
perturbation can be followed on its way through the atmosphere. We 
ran a test model with a sinusoidal source. In steady state the whole
atmosphere was varying sinusoidally, responding to the source.
A phase shift between the inner and outer
layers could also be observed.  All tests indicate that our extended
code works fine.

Future work is to calculate realistic light curves for Type Ia
supernova hydro models. As a first step we can consider
the atmosphere to be in LTE, but including detailed opacities and
treating full line blanketing.

Also with our
time-dependent radiative transfer, we can address a longstanding debate about
the importance of time dependence in calculating the spectra of Type
Ia supernovae \citep{eastpin93,bhm96}. One challenge is the
computation time needed for a whole light curve with more complex
NLTE radiative transfer.

\begin{acknowledgements} 
This work was supported in part by the 
Deutsche Forschungsgemeinschaft (DFG) via the SFB 676, NASA grant
NAG5-12127, NSF grant AST-0707704, and US DOE Grant DE-FG02-07ER41517.
This research used resources of the National Energy Research
Scientific Computing Center (NERSC), which is supported by the Office
of Science of the U.S.  Department of Energy under Contract
No. DE-AC02-05CH11231, and the H\"ochstleistungs Rechenzentrum Nord
(HLRN). We thank all these institutions for generous allocations of
computer time.
\end{acknowledgements}

\bibliographystyle{aa}
\bibliography{10982bib}

\end{document}